# Latency Reduction in CloudVR: Cloud Prediction, Edge Correction

Ali Majlesi Kopaee
Electrical Engineering Department
Sharif University of Technology
Tehran, Iran
majlesikopaee.ali@ee.sharif.edu

Seyed Amir Hajseyedtaghia
Electrical Engineering Department
Sharif University of Technology
Tehran, Iran
seyedamir.hajseyedtaghia@ee.sharif.edu

Hossein Chitsaz
Electrical Engineering Department
Sharif University of Technology
Tehran, Iran
hossein.chitsaz@ee.sharif.edu

Babak Hossein Khalaj*
Electrical Engineering Department
Sharif University of Technology
Tehran, Iran
khalaj@sharif.edu

**Abstract -** Current virtual reality (VR) headsets encounter a trade-off between high processing power and affordability. Consequently, offloading 3D rendering to remote servers helps reduce costs, battery usage, and headset weight. Maintaining network latency below 20ms is crucial to achieving this goal. Predicting future movement and prerendering are beneficial in meeting this tight latency bound. This paper proposes a method that utilizes the low-latency property of edge servers and the high resources available in cloud servers simultaneously to achieve cost-efficient, high-quality VR. In this method, head movement is predicted on the cloud server, and frames are rendered there and transmitted to the edge server. If the prediction error surpasses a threshold, the frame is re-rendered on the edge server. Results demonstrate that using this method, each edge server can efficiently serve up to 23 users concurrently, compared to a maximum of 5 users when rendering the frame entirely on the edge server. Furthermore, this paper shows that employing the Mean Absolute Error loss function and predicting acceleration rather than velocity significantly enhances prediction accuracy. Additionally, it is shown that normalizing individual data using its mean and standard deviation does not yield improvements in prediction accuracy. These findings provide insights into optimizing VR headset performance through cloud-edge collaboration.

## I. Introduction

Virtual reality headsets, essential for rendering 3D environments, demand substantial processing power. However, to alleviate costs, reduce weight, and minimize battery consumption, processing such as 3D rendering is being shifted outside the headset, relying on Cloud-based rendering and transmitting the imagery to the headset. Achieving minimal latency in this transmission is crucial. Studies by Hou et al. [1] and the Qualcomm Whitepaper [2] emphasize that for an optimal Virtual Reality experience, latency must remain under 20ms.

Maintaining a high-quality user experience within Cloud-based VR systems, especially in scenarios with higher latencies, poses a challenge. One proposed solution involves compensating for latency by predicting the future position of the user's head.

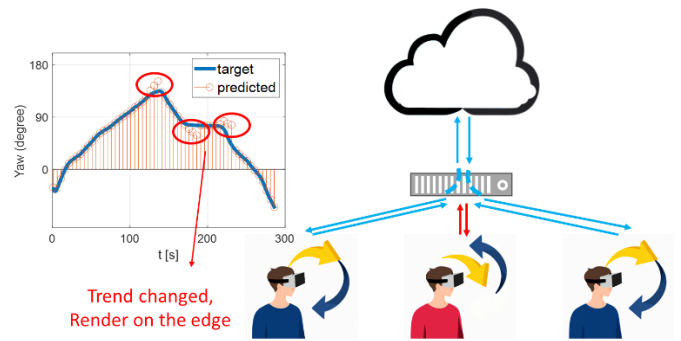

Fig. 1. Conceptual Overview of the Proposed Predictive Rendering Solution

## II. Literature Review

Research on head position prediction has primarily centered around two distinct environments:

1- 360-Degree Videos:

   Studies in this domain focus on predicting users' head angles while they watch 360-degree videos. In this context, only the user's head angles change, while their position remains fixed.

2- Virtual Reality Environments:

   Within virtual reality (VR) settings, users have the ability to move within the environment. Therefore, the predictive model needs to anticipate not only head angles but also head positions.

Gül et al. [3] employed a linear Kalman filter model, assuming constant angular head velocities for head position prediction.

Furthermore, Hou et al. [4] utilized an LSTM encoder-decoder architecture to predict head Euler angles. They predicted the x, y, and z coordinates by feeding each component into the LSTM encoder. The output of the three LSTMs was concatenated and fed into three fully connected decoders to predict each x, y, and z component.

In addition, Lee et al. [5] implemented an Attention model for predicting head positions, surpassing the accuracy of Recurrent Neural Networks.

Moreover, Yoon et al. [6] applied a constant velocity model to approximate head positions within an augmented reality (AR) environment. Subsequently, a neural network was employed to rectify errors generated by the constant velocity model.

Notably, prior studies did not explore the utilization of edge servers. However, this research introduces a novel methodology proposing the incorporation of edge servers to enhance prediction accuracy.

## III. Proposed Method

In this method, head positions and Euler angles are predicted on the cloud server, and frames are rendered there and sent to the edge server. Upon the edge server receiving both the predicted and real-time locations of the user, the edge server compares them. If the difference between these locations exceeds a certain threshold, the frame is re-rendered on the edge server; otherwise, the rendered frame from the cloud is forwarded to the user. This system is depicted in Fig. 1.

This paper is structured into two primary parts:

The first part focuses on employing a neural network encoder-decoder with an LSTM structure, similar to the one used by Hou et al. [4], for predicting head Euler angles in 360-Degree videos. However, unlike Hou et al. [4], this work does not predict head Euler angle velocities; instead, it utilizes head angular acceleration. This approach addresses an issue present when the model predicts speed without considering speed continuity. Predicting acceleration and subsequently deriving speed by integrating acceleration helps alleviate this problem. Furthermore, this section employs the Mean Absolute Error (MAE) loss function, diverging from the traditional Mean Square Error (MSE) loss function. This alteration prioritizes reducing errors in points with lower error rates more than those with higher error rates.

In the second part of this paper, the same neural network utilized in the first part is employed to predict users' head Euler angles. However, this section employs a dataset collected using MetaQuest2, where in addition to head Euler angles, it includes data on the user's head position while navigating within a VR space. This part uses an LSTM encoder-Fully connected decoder similar to the model employed by Hou et al. [4]. Unlike the data in the previous section, the intervals between consecutive data points vary from 10ms to 18ms in this dataset. This paper investigates the impact of inputting this time difference into the model on prediction accuracy.

Furthermore, an examination has been conducted to observe the effect on the model's accuracy when normalizing data using the mean and variance of each individual data rather than the mean and variance of the entire dataset.

### A. Head Euler angles prediction in 360-Degree Videos

For the prediction of head Euler angles, an LSTM encoder-decoder model similar to the one used by Hou et al. [4] was employed, as depicted in Fig 2. The model's hyperparameters are detailed in Table 1. Each component of the head Euler angles—$\phi$, $\theta$, $\psi$—was predicted separately using an encoder-decoder setup.

In the approach by Hou et al. [4], the model takes head angular velocities as input and predicts angular velocities as output. However, in this method, the model takes angular acceleration as input and predicts angular acceleration as output. This adjustment is based on the understanding that the head can be modeled as a rigid body influenced by muscle forces and torques. The motion of the head follows Newton's second law, which is a second-order ordinary differential equation. Muscles forces and torques serve as inputs, directly converted to acceleration. Therefore, it is expected that predicting acceleration will enhance accuracy.

Moreover, predicting angular speeds often leads to discontinuities in angular speed continuity. However, by predicting acceleration and subsequently deriving speeds through integration, these discontinuities are mitigated.

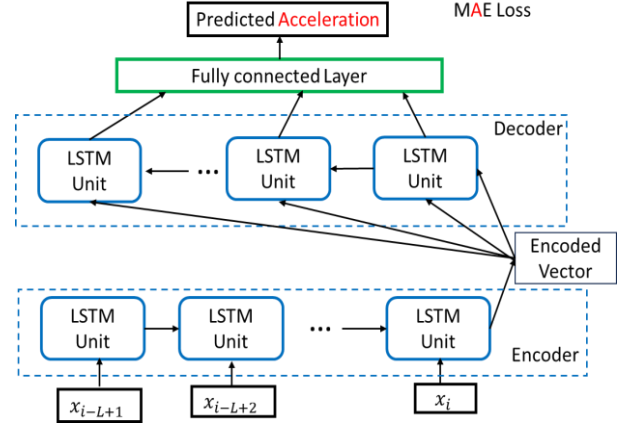

Fig. 2. Neural Network Structure for Predicting User's Head Angles

TABLE I. HYPERPARAMETERS OF THE NEURAL NETWORK

|  | **Meta Quest2 Data** | **360-Degree Video Data** |
|---|---|---|
| Hidden LSTM Layer Dimensions | 30 | 30 |
| Input Window Length | 60 samples | 30 samples |
| Time Interval Between the Last Data and the Predicted Data | 60ms | 60ms |
| Time Interval Between Samples | 10ms in average | 30ms |

TABLE II. MODEL ACCURACY IN TWO STATES OF VELOCITY PREDICTION AND ACCELERATION PREDICTION, AND WITH MAE AND MSE LOSS FUNCTION.

| Predicted Parameter | Velocity [4] | Velocity | Acceleration | Acceleration |
|---|---|---|---|---|
| Loss Function | MSE | MAE | MSE | MAE |
| MAE [degree] | 0.630 | 0.715 | 0.440 | 0.391 |
| MAE with limiting error to less than 1 degree [degree] | 0.332 | 0.382 | 0.259 | 0.227 |
| Percentage of points with an error of more than 1 degree | 11.04% | 14.54% | 7.51% | 5.79% |
| MSE [degree] | 2.109 | 2.139 | 1.435 | 1.449 |

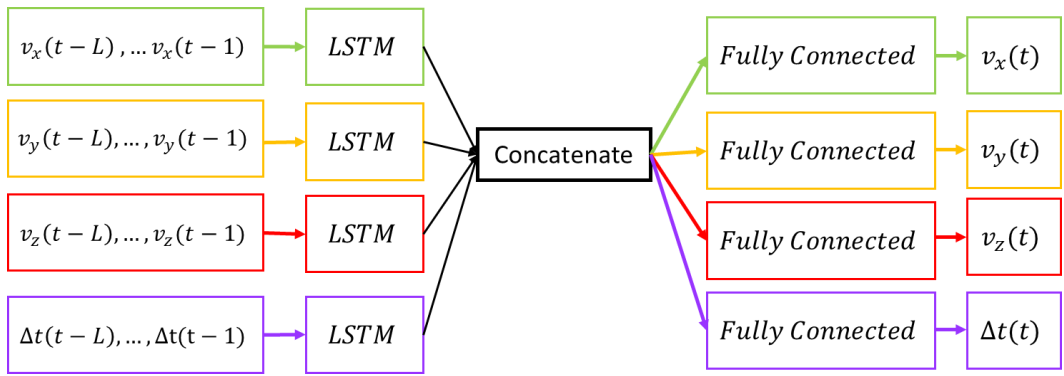

Fig. 3. Neural Network Structure for Predicting User's Head Position

*1) Comparison of Acceleration Prediction and Velocity Prediction*

The prediction accuracy for two different configurations—speed prediction and acceleration prediction—is shown in Table 2. The percentage of error exceeding 1 degree in acceleration prediction has decreased to 5.87% from 12.8% observed in speed prediction.

*2) Comparison of MAE and MSE Loss Functions*

For the evaluation metric, Mean Absolute Error (MAE) was employed by clipping errors exceeding 1 degree. This choice was made because when errors surpass 1 degree, the frame is re-rendered on the edge server to accommodate the user's online position. The metrics are displayed in Table 2 under 'MAE with limiting error to less than 1 degree'. Additionally, 'Percentage of points with an error of more than 1 degree' represents the percentage of points where the prediction error exceeded 1 degree.

The utilization of MAE loss results in a decrease in the "MAE with limiting error to less than 1 degree' metric but an increase in the 'Percentage of points with an error of more than 1 degree'. This occurrence is due to the effect of employing the MAE loss function. Minimizing the MAE loss prioritizes reducing errors in points where the model predicts more confidently than in other points [8]. Consequently, in this method, points with higher errors are re-rendered on the edge server, emphasizing their reduced error importance.

Bassbouss et al. [9] observed that when more than 5 users are served on a single T4 GPU, the rendering frame rate drops below 37 fps. If the frame rate falls below 30 fps, users experience delays between frames. The number of users that can be simultaneously served on a single edge server using this method can be calculated by determining the probability of misprediction when more than 5 users are present. This calculation can be expressed as follows:

$$P_{5users} = \sum_{k=5}^{m} \binom{m}{k} P_{error}^{k} (1 - P_{error})^{m-k} \quad (1)$$

Where $P_{error} = 0.0579$, and the threshold is 1%, the condition 'the probability of more than 5 users is below a threshold' translates to $P_{5users} < 0.01$, indicating that m should be limited to m ⩽ 23. Thus, an edge server can simultaneously service up to 23 users.

*B. Head orientation and position prediction in a VR Environment*

The neural network depicted in Fig. 3 has been employed for head position prediction. In this model, the past L samples the $v_x, v_y, v_z$ and dt components are separately fed into LSTM networks. The outputs of these four LSTM networks are concatenated and then fed into four fully connected networks to generate $v_x, v_y, v_z$ and dt values for the next frame.

This model was evaluated in four different configurations:

In two configurations, each stream of data in which users explored the VR environment for 1 to 2 minutes was normalized using the average and standard deviation (STD) of that one-minute stream. It was anticipated that this method would enhance the model's generalization and accuracy. These models are labeled as 'stream normalization' in Table 3.

In the other two configurations, the entire dataset was normalized using the average and STD. These models are indicated as 'normalization' in Table 3. However, the results indicate that 'stream normalization' does not improve model accuracy. Nevertheless, it is worth noting that longer durations for each data stream might yield improved results. To evaluate the effect of incorporating Δ𝑡 as an input, two configurations (as shown in Fig. 3) involved feeding Δ𝑡 to the model. These configurations are labeled as '+dt' in Table 3. In the other two configurations, Δ𝑡 was not included as an input to the model. The results indicate that providing Δ𝑡 as an input to the model does not significantly improve accuracy.

TABLE III. COMPARISON OF FOUR STATES WITH ΔT INPUT AND WITHOUT IT, USING STREAM NORMALIZATION AND WITHOUT IT

| Method | dx error | dy error | dz error | dt error [ms] |
|---|---|---|---|---|
| Average of parameter [mm] | 15.0 | 3.7 | 15.1 | 70.7 |
| stream normalization + dt [mm] | 3.27 | 1.80 | 3.32 | 2.40 |
| normalization + dt [mm] | 3.02 | 1.73 | 3.14 | 1.3 |
| stream normalization [mm] | 3.20 | 1.79 | 3.30 | _ |
| normalization [mm] | 3.03 | 1.74 | 3.17 | _ |

## IV. Dataset

In section III-A, the 'Full UHD 360-Degree Video' [7] dataset was utilized. These data comprise the head Euler angles of users while watching 360-degree videos. This dataset contains data from 34 users, with each data stream lasting approximately one to two minutes, totaling 190 streams of data. The data were collected at a sample rate of 30 samples per second.

In section III-B, data were collected from users exploring a VR environment using the MetaQuest 2 headset. This dataset comprises information from 25 users and includes 58 one to two-minute streams of data. Seven VR environments were utilized to collect this dataset; however, not all seven environments were presented to each user.

## V. Conclusion

This paper proposes a method for distributing rendering between edge and cloud computing. It demonstrates that predicting acceleration and integrating it to calculate speed enhance accuracy when compared to solely predicting speed in 360-degree videos. Additionally, the evaluation presented in this paper shows that using Mean Absolute Error (MAE) loss, rather than Mean Squared Error (MSE) loss, improves model accuracy.

Furthermore, the evaluation conducted in this paper indicates that including $\Delta t$ in variable sample rate scenarios as input to the model and performing stream normalization does not lead to an improvement in model accuracy.

## VI. Future Work

It is worthwhile to assess the methods presented in this paper using a neural network with an attention mechanism. Additionally, comparing previous models on a single dataset would be beneficial for evaluating their relative performance.

Another valuable pursuit is to approximate the maximum achievable accuracy for head position prediction. To achieve this, an experiment can be conducted where users are constrained to follow a specified path for 3 to 4 seconds before reaching a predefined point in space. This approach aims to unify the history of head movements. By training a model on this data, the resulting error can serve as a lower bound for the potential achievable accuracy.